\documentstyle[emulateapj,apjfonts,psfig]{article}

\def\refnew#1{(\ref{#1})}
\def\be{\begin{equation}}
\def\ee{\end{equation}}

\def\vcv{v_{\rm cv}}
\def\tcv{t_{\rm cv}}

\def\cs{c_s}

\def\K{\, \rm K}

\def\be{\begin{equation}}
\def\ee{\end{equation}}

\def\bx{\bf x}
\def\bk{\bf k}

\begin{document} 

\submitted{Submitted to ApJ on Oct. 2nd, 1998}

\title{\mbox{GRAVITY-MODES IN ZZ CETI STARS} \\ \mbox II.
EFFECTS OF TURBULENT DISSIPATION}
\author{Peter Goldreich\altaffilmark{1} and Yanqin Wu\altaffilmark{1,2}}

\altaffiltext{1}{Theoretical Astrophysics, California Institute of Technology,
	130-33, Pasadena, CA 91125, USA; pmg@gps.caltech.edu}
\altaffiltext{2}{Astronomy Unit, School of Mathematical Sciences, 
	 Queen Mary and Westfield College, Mile End Road, London E1 4NS, UK;
                 Y.Wu@qmw.ac.uk}

\begin{abstract}
We investigate dynamical interactions between turbulent convection and
g-mode pulsations in ZZ Ceti variables (DAVs). Since our understanding
of turbulence is rudimentary, we are compelled to settle for order of
magnitude results. A key feature of these interactions is that
convective response times are much shorter than pulsation periods.
Thus the dynamical interactions enforce near uniform horizontal
velocity inside the convection zone. They also give rise to a narrow
shear layer in the region of convective overshoot at the top of the
radiative interior. Turbulent damping inside the convection zone is
negligible for all modes, but that in the region of convective
overshoot may be significant for a few long period modes near the red
edge of the instability strip. These conclusions are in accord with
those reached earlier by Brickhill. Our major new result concerns
nonlinear damping arising from the Kelvin-Helmholtz instability of the
aforementioned shear layer.  Amplitudes of overstable modes saturate
where dissipation due to this instability balances excitation by
convective driving. This mechanism of amplitude saturation is most
effective for long period modes, and it may play an important role in
defining the red edge of the instability strip.
\end{abstract}


\setcounter{equation}{0}

\section{Introduction
\label{sec:na-vis-intro}}

The current paper focuses on dynamical interactions between turbulent 
convection and g-mode pulsations in ZZ Ceti variables. It is the second of a 
series, and the last in which we rely on order of magnitude reasoning and the 
quasiadiabatic approximation. The reader is referred to Goldreich \& Wu 
\cite{vis-paper1} (hereafter Paper I) for descriptions 
of the properties of g-modes and scaling relations appropriate to turbulent 
convection. Only results of immediate relevance to the current investigation are 
quoted here. Symbols taken from Paper I are defined in Table 1.

\begin{deluxetable}{cll}
\tablewidth{0pc}
\tablecaption{DEFINITIONS}
\tablehead{
\colhead{Symbol}      & \colhead{Meaning}} 
\startdata
$R$ & stellar radius \nl
$g$ & surface gravity \nl
$r$ & radial distance from center of star \nl
$z$ & depth below photosphere \nl
$z_b$ & depth at bottom of convection zone \nl
$z_\omega$ & depth at top of mode's cavity,\, $z_\omega\sim \omega^2/(gk_h^2)$ 
\nl
$\omega$ & radian mode frequency \nl
$n$ & radial order of mode \nl
$\ell$ & angular degree of mode \nl
$k_h$ & horizontal wave vector,\,  $k_h^2=\ell(\ell+1)/R^2$ \nl
$\rho$ & mass density \nl
$p$ & pressure \nl
$s$ & specific entropy in units of $k_B/m_p$ \nl
$F$ & energy flux \nl
$L$ & luminosity,\,  $L=4\pi R^2F$  \nl
$c_s$ & adiabatic sound speed,\,  $c_s^2=(\partial p/\partial \rho)_s$ \nl
$\rho_s$ & $(\partial\ln\rho/\partial s)_p$ \nl
$\delta$ & denotes Lagrangian perturbation \nl
$\xi_h$ & horizontal component of displacement vector \nl
$\xi_z$ & vertical component of displacement vector \nl
$\vcv$ & convective velocity,\,  $\vcv\sim (F/\rho)^{1/3}$ \nl
$\tcv$ & response time for convection,\,  $\tcv\sim z/\vcv$ \nl
$\nu$ & turbulent kinematic viscosity,\,  $\nu\sim z\vcv$ \nl
$A, \,\, B, \,\, C$ & dimensionless constants approximately $2, \,\, 8$ \& $8$
for ZZ Cetis \nl
$\tau_{\rm th}$ & thermal constant at depth z,\, $\tcv/\tau_{\rm th}\sim
(\vcv/c_s)^2$ in the convection zone \nl
$\tau_b$ & unconventional thermal time constant at $z_b$,\, $\tau_b\approx 
\tau_{\rm th}/5$ at $z_b$.  \nl
$\tau_c$ & time constant of low pass filter for convection zone, \, 
$\tau_c=(B+C)\tau_b$ \nl
\enddata
\end{deluxetable}

ZZ Ceti variables possess surface convection zones. Thus the
interaction of their g-modes with turbulent convection requires
attention. An important feature of convection in these stars is that
its response time is much shorter than the periods of the observed
modes. Thus to a good approximation, the convection adjusts to the
instantaneous pulsational state. This behavior plays an essential role
in the overstability mechanism (Brickhill \cite{vis-brick90}, Gautschy
\cite{vis-gautschy96}, Paper I)
which, following Brickhill, we refer to as convective driving.
 
The nonlinear advection term in the fluid momentum equation describes
interactions between g-modes and turbulence.  We model these
interactions as giving rise to a turbulent viscosity which acts to
reduce the magnitude of a mode's shear tensor.  We devote \S 2 to
evaluating the effects of turbulent viscosity on g-modes, both inside
the convection zone and in the region of convective overshoot at the
top of the radiative interior.  We apply these results in \S 3 to
estimate the contributions of turbulent viscosity to mode damping.  In
\S 4, we analyze the saturation of the amplitude of an overstable mode
due to instability of the narrow shear layer it forces at the top of
the radiative interior. We conclude with a brief discussion in \S 5.

\section{Effects Of Turbulent Viscosity
\label{sec:na-vis-reduct}}

Since motion in the outer layers of a DA white dwarf is all that
concerns us, it proves convenient to adopt a plane-parallel
approximation. Perturbation theory is done with Lagrangian variables
which are assumed to carry a time dependence $e^{-i\omega t}$ and a
horizontal spatial dependence $e^{i\bk_h\cdot\bx}$. Gravity modes
involve a mostly horizontal sloshing of fluid. Our principal concern is
with the vertical derivative of the horizontal component of the
displacement, $d\xi_h/dz$, because it is the largest component of the
gradient of the displacement.

In \S \ref{subsec:na-vis-reducta} we calculate the magnitude of 
$d\xi_h/dz$ inside the convection zone. To begin, we estimate the magnitude
that $d\xi_h/dz$ would have in the absence of turbulent viscosity. In the dual 
limit of an isentropic convection zone and a vanishing entropy perturbation, 
$|d \xi_h /dz|$ would be equal to $\ell |\xi_z|/R$ which is smaller than $\ell 
|\xi_h|/R$. However, a real convection zone is not an isentrope; its specific 
entropy increases with depth. Moreover, a g-mode produces a specific entropy 
perturbation. It turns out that the latter departure from isentropy is more 
important than the former in enhancing the magnitude of $d\xi_h/dz$. Without 
turbulent viscosity, the entropy perturbation could induce a $|d\xi_h/dz|$ as 
large as $\xi_h/z_\omega$. Inclusion of turbulent viscosity reduces $|d 
\xi_h/dz|$ by a multiplicative factor $\omega \tcv\ll 1$.

In \S \ref{subsec:na-vis-reductb} we examine the velocity shear in
the transition region between the base of the convection zone and the top
of the radiative interior. The turbulent viscosity declines sharply
with depth in this region. We approximate this decline by a
discontinuous drop to zero viscosity.  The application of appropriate
boundary conditions then shows that the horizontal velocity is
discontinuous at $z_b$. We evaluate the magnitude of this velocity
jump and the associated jump in the gradient of the pressure
perturbation.

\subsection{Velocity Shear in the Convection Zone
\label{subsec:na-vis-reducta}}

The linearized equations of mass and momentum conservation, the latter
in component form, read (cf. Paper I)
\begin{eqnarray}
{\delta\rho\over\rho} & =& -ik_h\xi_h-{d\xi_z\over dz},
\label{eq:na-vis-cont} \\
\omega^2 \xi_h&=&i k_h \left({p\over\rho} {{\delta p}\over p} - g
\xi_z \right) -{f_h\over \rho},
\label{eq:na-vis-xh} \\
\omega^2 \xi_z&=&{p\over\rho} {d\over{dz}}\left({{\delta p}\over
p}\right) + g\left({{\delta p}\over{p}} + i k_h \xi_h \right) -{f_z\over \rho},
\label{eq:na-vis-xz}
\end{eqnarray}
where $f_i$ is the force per unit volume due to turbulent convection.

For the moment, we neglect the force due to turbulence. Then differentiating 
equation \refnew{eq:na-vis-xh} and substituting for $d\xi_z/dz$ using equation 
\refnew{eq:na-vis-cont}, we obtain
\be
{d\xi_h\over dz}={i k_h\over \omega^2}
\left[{d\over dz}\left({p\over\rho}{\delta p\over p}\right) + g\left(
{\delta \rho\over \rho} + i k_h \xi_h \right) \right].
\label{eq:na-vis-diffxi}
\ee
With the aid of equation \refnew{eq:na-vis-xz} and the equation of state,
\be
{\delta\rho\over\rho}={p\over c_s^2\rho}{\delta p\over p}+
\rho_s \delta s, \label{eq:adia-nonadden}
\ee
we recast equation \refnew{eq:na-vis-diffxi} in a more revealing form
\be
{d\xi_h\over dz}=ik_h\xi_z-{igk_h\rho_s\over\omega^2}\left[
{p\over{g\rho}}{ds\over dz}
\left({\delta p\over p}\right) - \delta s \right].
\label{eq:adia-shearp} 
\ee

We can think of the $\delta p/p$ and $\delta s$ terms as providing
adiabatic and nonadiabatic forcing of $d\xi_h/dz$.  Both terms vanish
for adiabatic perturbations in an isentropic convection zone.  In
their absence, the perturbations are irrotational so
$d\xi_h/dz=ik_h\xi_z$.\footnote{This is a linearized version of
Kelvin's circulation theorem.} This value of shear would lead to
negligible turbulent damping.  To estimate $d\xi_h/dz$ when the flow
is rotational, we must relate $\delta p/p$ and $\delta s$ to
$\xi_h$. Here we appeal to Paper I which establishes that for
$z\lesssim z_b\lesssim z_\omega$,\footnote{Overstable modes have
$z_\omega>z_b$. These are the only ones that concern us in this
paper.}
\be
{\delta p\over p}\approx -ik_h\xi_h,
\label{eq:na-vis-dpxhequiv}
\ee
and
\be
\delta s\approx {A(B+C)\over 1-i\omega\tau_c}\left({\delta p\over p}\right).
\label{eq:na-vis-dels}
\ee
The ratio of adiabatic to nonadiabatic forcing, as estimated at $z =
z_b$, is given by
\begin{eqnarray}
\left({p\over g\rho}{ds\over dz}\right)_b \biggl\vert{{\delta p}\over
p}\biggr\vert_b {1\over |\delta s|_b} & \sim &  \left({p\over g\rho}{ds\over 
dz}\right)_b{[1+(\omega\tau_c)^2]^{1/2}\over A(B+C)}
\nonumber \\
& \sim &  \left[\left({\vcv\over 
\cs}\right)^4+\left(\omega\tcv\right)^2\right]_b^{1/2}\ll 1,
\label{eq:adia-ratio}
\end{eqnarray}
where we make use of the relations $gds/dz \sim (v_{\rm cv}/c_s)^2$
and $\tcv/\tau_b\sim (\vcv/c_s)^2$. This establishes nonadiabatic
forcing as the principal driver of velocity shear in the convection
zone. In an inviscid convection zone, $|d\ln\xi_h/dz|$ would be of
order $z_\omega^{-1}$. Acting on a shear of this magnitude,
turbulent damping would overwhelm convective driving.\footnote{A similar 
conclusion is arrived at by Brickhill (\cite{vis-brick90}).} 

Next, we assess the effects of turbulence on the shear. Given our present
understanding, about the best we can do is to model the turbulent Reynolds 
stress by analogy with the stress due to molecular viscosity in a Newtonian 
fluid. The viscous force per unit volume is given by
\be
f_i={\partial\sigma_{ij}\over\partial x^j}, 
\label{eq:na-vis-f}
\ee
where the stress tensor, $\sigma_{ij}$, is related to the g-mode's
displacement field by two scalar coefficients, the shear viscosity,
$\nu$, and the bulk viscosity, $\zeta$, according to
\be
\sigma_{ij}=-i\omega\rho\left\{\nu\left[\left({\partial \xi_i\over\partial 
x^j}+{\partial \xi_j\over\partial x^i}\right) - {2\over 3}{\partial 
\xi_k\over\partial x^k}\delta_{ij}\right]+\zeta {\partial \xi_k\over\partial 
x^k}\delta_{ij}\right\}.
\label{eq:na-vis-sigma}
\ee
When the effects of turbulent viscosity are neglected, $d\xi_h/dz$ is
by far the largest component of the gradient of the
displacement. Since this term contributes to the shear but not to the
divergence, we need not consider terms involving the bulk
viscosity. Also, it is easy to show that $f_z/g\rho$ is much smaller
than either $\delta p/p$ or $ik_h\xi_h$. Therefore, we ignore the
effects of turbulent viscosity in the vertical component of the
momentum equation.

Turbulent viscosity has a profound effect on the horizontal component of the 
momentum equation. The dominant term in $f_h$ takes the form\footnote{We neglect 
the depth variation of $\xi_h$ with respect to that of $\rho\nu$.}
\be
f_h\approx -i\omega{d\over dz}\left(\rho\nu{d\xi_h\over dz}\right)\sim 
-i\omega{\rho\vcv}{d\xi_h\over dz}.
\label{eq:na-vis-alphah}
\ee
Adding the approximate expression for $-f_h/\rho$ to the right hand side of 
equation \refnew{eq:na-vis-xh} yields
\be
\omega^2 \xi_h\approx ik_h \left({p\over\rho} {{\delta p}\over p} - g
\xi_z \right) + {i\omega z\over \tcv}{d\xi_h\over dz}.
\label{eq:na-vis-xhvis} 
\ee
Differentiating this equation with respect to $z$, and repeating the
steps that led to equation \refnew{eq:adia-shearp}, we obtain
\be
{d\xi_h\over dz}\sim {\omega\tcv\over 1-i\omega\tcv}\left\{ 
k_h\xi_z-{gk_h\rho_s\over\omega^2}\left[ {p\over{g\rho}}{ds\over dz}
\left({\delta p\over p}\right) - \delta s \right]\right\}.
\label{eq:adia-shearpvis} 
\ee
Since $\omega\tcv\ll 1$, turbulent viscosity significantly reduces
$d\xi_h/dz$. 

For completeness, we provide an estimate for $d\xi_z/dz$. Combining equations 
\refnew{eq:na-vis-cont}, \refnew{eq:adia-nonadden}, 
\refnew{eq:na-vis-dpxhequiv}, and \refnew{eq:na-vis-dels} yields 
\be
{d\xi_z\over dz}\sim -ik_h\xi_h,
\label{eq:na-vis-dvzdz}
\ee
which is independent of the magnitude of the turbulent viscosity. The
small value of $d\xi_z/dz$ ensures that its contribution to mode
damping is negligible, as are the contributions from $ik_h\xi_h$ and
$ik_h\xi_z$.

\subsection{Shear Layer at the Base of the Convection Zone
\label{subsec:na-vis-reductb}}

Here we show that there is a shear layer at the top of the radiative
interior across which the horizontal displacement jumps by an amount
similar to the depth integrated change it would undergo in an inviscid
convection zone.

We assume that the turbulent viscosity drops discontinuously to zero across the 
boundary at $z=z_b$. Then the continuity of the tangential stress demands 
that\footnote{We denote by $b^\mp$ quantities evaluated on the 
convective and radiative sides of the boundary.} 
\be
{d\xi_h\over dz}\bigg|_{b^-}=0.
\label{na-vis-dxihdzb}
\ee
Next we multiply equation \refnew{eq:na-vis-xhvis} by $\rho$ and integrate
the resulting expression from the top of the stellar atmosphere to the
bottom of the convection zone. This procedure yields
\be
\omega^2 \int_0^{z_b} dz\, \rho\, \xi_h \approx i k_h \int_0^{z_b} dz\, \rho 
\left[ {p\over\rho} \left({{\delta p}\over p}\right) - g \xi_z \right].
\label{eq:na-vis-integxhcvz}
\ee  
We transform equation \refnew{eq:na-vis-integxhcvz} in two steps. Integrating
by parts, we obtain
\begin{eqnarray}
\omega^2 \int_0^{z_b} dz\, \rho\, \xi_h & \approx  &  
  {ik_hp_b\over g}\left[ 
{p\over\rho} \left({{\delta p}\over p}\right) - g \xi_z
\right]_{b^-} \nonumber \\ 
& & - {ik_h\over g}\int_0^{z_b} dz\, p {d\over dz}
\left[ {p\over\rho} \left({{\delta p}\over p}\right) - g \xi_z \right].
\label{eq:na-vis-identity} 
\end{eqnarray}
Then using equations \refnew{eq:na-vis-cont}, \refnew{eq:na-vis-xz}, and 
\refnew{eq:adia-nonadden}, and taking into account the continuity of $\delta p$ 
and $\xi_z$ across $z_b$, we arrive at\footnote{We drop the negligible 
$\omega^2\xi_z$ and $-f_z/\rho$ terms in equation \refnew{eq:na-vis-xz}.}
\be  
\omega^2 \int_0^{z_b} dz\, \rho\, \xi_h \approx {\omega^2 p_b\over g} 
\xi_h|_{b^+}\,\, +\,\, ik_h\int_0^{z_b} dz\, p\rho_s\left[{p\over g\rho} 
{ds\over dz} 
\left({\delta p\over p}\right)-\delta s\right].
\label{eq:na-vis-transf}
\ee
Since for $\omega\tcv\ll 1$, $\xi_h$ is nearly constant within the convection 
zone, we establish the discontinuity of $\xi_h$ across $z_b$ to be 
\be
\Delta\xi_h\equiv\xi_h\biggr\vert_{b^-}^{b^+} \approx {{-igk_h}\over{\omega^2 
p_b }} 
\int_0^{z_b} dz\, p\rho_s
\left[{p\over g\rho}{ds\over dz}\left({{\delta p}\over p}\right) -
\delta s  \right].
\label{eq:na-vis-expressjump}
\ee
It follows directly from equation \refnew{eq:na-vis-xz} that
\be
{p\over g\rho}{d\over dz} \left({{\delta p}\over
p}\right)\biggr\vert^{b^+}_{b^-}
\approx  - i k_h \xi_h\biggr\vert^{b^+}_{b^-}.
\label{eq:na-vis-jumponddp}
\ee

Equation \refnew{eq:na-vis-expressjump} merits a few comments. It may
be simplified by taking advantage of the near constancy of $\delta
p/p$ and $\delta s$ inside the convection zone. Both are due to rapid
mixing by turbulent convection which occurs on the time-scale $\tcv$. 
Since the term
involving the entropy perturbation dominates that due to the
unperturbed entropy gradient, and $|\delta s|\sim |\delta p/p|\sim
|k_h\xi_h|$, the fractional change in $\xi_h$ across $z_b$ is of order
$z_b/z_\omega$. Thus the relative size of the jump is greatest for
long period modes in cool stars. Finally, as advertised, comparison
with equation \refnew{eq:adia-shearp} reveals that the discontinuity
in $\xi_h$ is closely related to the total variation that $\xi_h$
would experience within the convection zone in the absence of
turbulent viscosity.

\section{Turbulent Damping
\label{sec:na-vis-esti}}

Damping due to turbulent viscosity is evaluated inside the convection zone in \S 
\ref{subsec:na-vis-estia}, and in the region of convective overshoot in \S 
\ref{subsec:na-vis-estia}.\footnote{Here we treat the region of convective 
overshoot as having a finite extent unlike what we did in \S
\ref{subsec:na-vis-reductb}.} The former is shown to be negligible for all 
modes.
However, the latter may stabilize low frequency modes which would otherwise be 
overstable. 

\subsection{Inside the Convection Zone
\label{subsec:na-vis-estia}}

Damping due to turbulent viscosity within the convection zone may be
estimated as
\be
\gamma_{\rm vis} \approx -{\omega^2R^2\over 2}\int_0^{z_b}\, dz\, \rho\,
\nu \left({d\xi_h\over dz}\right)^2, \label{eq:na-vis-Edot}.
\ee
Evaluating this integral with only the (dominant) entropy perturbation
term from equation \refnew{eq:na-vis-dels} retained on the right hand
side of equation \refnew{eq:adia-shearpvis}, we obtain
\begin{equation}
\gamma_{\rm vis-cv} \sim -0.1{\omega^2\tcv\tau_c\over
1+(\omega\tau_c)^2}\left({z_b\over z_\omega}\right)L \left({\delta p\over 
p}\right)_b^2,
\label{eq:adia-Wdotviscp}
\end{equation}
where $(\delta p/p)_b\sim 1/(n \tau_\omega)$ is the normalized
eigenfunction evaluated at $z_b$.  In deriving equation
\refnew{eq:adia-Wdotviscp}, we integrate over an isentropic convection
zone with adiabatic index $5/3$, and substitute for $A$, $B$, $C$,
$\vcv$, and $\tcv$ according to Table 1. Radiative damping as
evaluated in Paper I yields
\be
\gamma_{\rm rad}\sim -0.01L\left({\delta p\over p}\right)_b^2. 
\label{eq:adia-Wdotr}
\ee
Since $\tcv/\tau_c\ll 0.1$ for DA white dwarfs crossing the ZZ Ceti
instability strip, turbulent damping inside the convection zone is
much smaller than radiative damping for all overstable g-modes.

It follows from equations \refnew{eq:adia-shearpvis} and
\refnew{eq:na-vis-Edot} that turbulent damping of a given mode is
maximized for $\omega\tcv\sim 1$; it is negligible inside the
convection zone because $\omega\tcv\ll 1$ there.\footnote{It might
seem paradoxical, but provided $\omega\tcv\ll 1$, turbulent damping is
inversely proportional to the magnitude of the turbulent
viscosity. This follows because the dissipation rate is proportional
to $(z^2/\tcv)(d\xi_h/dz)^2$, and $|d\xi_h/dz|\propto
\omega\tcv$ for $\omega\tcv\ll 1$.} However, $\omega\tcv$ rises with 
depth in the region of convective overshoot. This suggests that
damping in this region may exceed that inside the convection zone.

\subsection{In the Region Of Convective Overshoot
\label{subsec:na-vis-estib}}

The intensity of turbulence in this region diminishes with depth. We
estimate its contribution to mode damping by solving a model
problem. We replace the bottom of the convection zone by a horizontal
plate, and the region of convective overshoot by an underlying viscous
fluid of uniform density whose viscosity decays exponentially with
depth,
\be
\nu (z) = \nu_b e^{-(z-z_b)/\lambda}.
\label{eq:vis-nuovershoot}
\ee
The distance below the convection zone over which turbulent mixing
maintains the entropy gradient small and negative is of order
$\lambda\ln(c_s/\vcv)_b^2$, or several times as large as $\lambda$.

The shear layer is driven by the plate which oscillates at frequency
$\omega$ with displacement $-\Delta\xi_h$ given by
\refnew{eq:na-vis-expressjump}.  The Navier-Stokes equation yields
\be
{d\over{dz}}\left(\nu{{d \xi_h}\over{dz}}\right)- i\omega\xi_h=0,
\label{eq:vis-xihovershoot}
\ee
which is to be solved subject to the boundary conditions $\xi_h = - \Delta 
\xi_h$ at $z = z_b$, and $\xi_h \to 0$ as $z \to \infty$. Next we
define 
\be
\epsilon=\left({\omega\lambda^2\over \nu_b}\right)^{1/2} \ll 1,
\label{eq:vis-defeps}
\ee
change independent variable to 
\be
y= 2\epsilon e^{(z-z_b)/2\lambda-i\pi/4},
\label{eq:vis-defsz}
\ee
and dependent variable to 
\be
u= e^{-(z-z_b)/2\lambda}\xi_h.
\label{eq:vis-defu}
\ee
With these changes, equation \refnew{eq:vis-xihovershoot} is recast in
a more familiar form
\be
{d^2 u\over dy^2}+{1\over y}{du\over dy}+\left(1-{1\over y^2}\right)u= 0,
\label{eq:vis-ostrans3}
\ee
which is Bessel's equation for $n=1$. The appropriate solution, which
vanishes as $|y|\to \infty$, is $H_1^{(2)}=J_1-iY_1$, where $J_1$,
$Y_1$ and $H_1^{(2)}$ are the Bessel functions of the first kind, the
second kind and the third kind, respectively (cf. Abramowitz \& Stegun
\cite{abramowitz70}). 
The solution for $\xi_h$ may be written as
\be
\xi_h=DyH_1^{(2)}(y), 
\label{eq:vis-solnA}
\ee
where the coefficient $D$ is evaluated by applying the boundary condition
$\xi_h=-\Delta\xi_h$ at $z=z_b$. Since $|y_b|=2\epsilon\ll 1$, we adopt the
approximation $y_bH_1^{(2)}(y_b)\approx 2i/\pi$, which implies
\be
D\approx {i\pi\Delta\xi_h\over 2}.
\label{eq:vis-valD}
\ee

The viscous stress the plate exerts on the fluid is given by
\be
S=i\omega\rho_b\nu_b{d\xi_h\over dz}.
\label{eq:vis-S}
\ee
Making use of the identity
\be
{d\over dy}(yH_1^{(2)})=yH_0^{(2)}, 
\ee
we find
\be
S={-\pi\omega\rho_b\nu_b\over 4\lambda}\Delta\xi_h y_b^2H_0^{(2)}(y_b)\approx
2\omega^2\lambda\rho_b\Delta\xi_h\left(\ln\epsilon+\gamma+i\pi/4\right),
\label{eq:vis-Sexpl}
\ee
where $\gamma$ is Euler's constant. 

\begin{figure*}[t]
\centerline{\psfig{figure=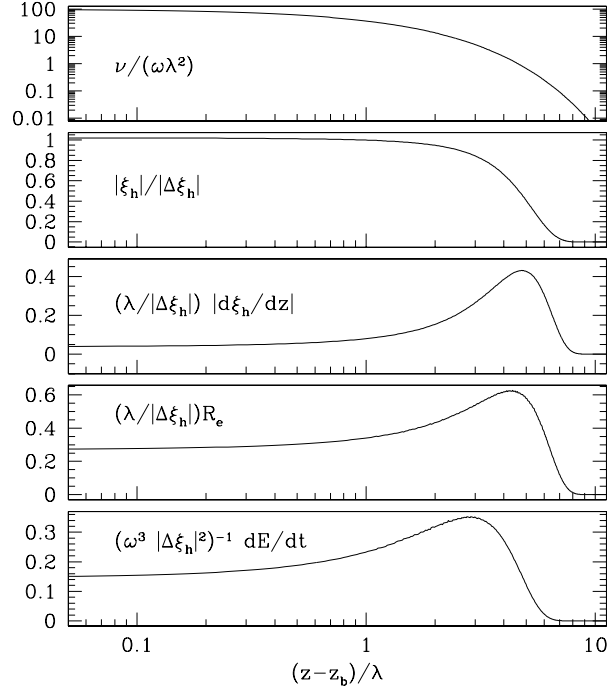,width=0.55\hsize}}
\caption[]{Numerical solution to the toy model in 
\S \ref{subsec:na-vis-estia} for $\epsilon=0.1$. 
From top to bottom the panels display scaled dimensionless versions of
the viscosity, $\nu$, the magnitude of the horizontal displacement,
$|\xi_h|$, the magnitude of the principal component of shear,
$\omega|d\xi_h/dz|$, the local Reynolds number, $Re\equiv
(\omega|\xi_h|^2)/(\nu|d\xi_h/dz|)$, and the viscous rate of energy
dissipation per unit mass, $dE/dt\equiv
\nu\omega^2|d\xi_h/dz|^2$, as functions of distance below $z_b$
measured in units of $\lambda$. Each of the curves plotted in the
bottom four panels is a function of the single variable
$\nu/(\omega\lambda^2)$ and the parameter $\epsilon$. Moreover, the
$\epsilon$ dependence is weak (logarithmic) for $\epsilon\ll 1$. Note
that the shear, the Reynolds number, and the rate of energy
dissipation all peak near the depth where$\nu/(\omega\lambda^2) =1$.}
\label{fig:na-vis-plate2}
\end{figure*}

The power per unit area supplied to the fluid, $P/A$, is equal to the product 
of $S$ with the plate's velocity, $i\omega\Delta\xi_h$. To obtain the time 
averaged rate of energy dissipation, we multiply $P/A$ by $4\pi R^2$ and average
over both time and solid angle to arrive at
\be
\left\langle{dE\over dt}\right\rangle\approx {\pi R^2\rho_b\omega^3\lambda| 
\Delta\xi_h|^2\over 4}.
\label{eq:vis-dEdt}
\ee
Note that $\langle{dE/dt}\rangle$ is independent of $\nu_b$ and
proportional to $\lambda$. This is not surprising. Close to the plate the
fluid moves almost rigidly, so the rate of dissipation per unit volume is small,
just as it is in the interior of the convection zone. This rate first increases 
with depth, reaching a peak where the viscous diffusion time is of order 
$\omega^{-1}$, and subsequently declines. Near the peak, $\nu\sim 
\omega\lambda^2$, which implies a dissipation rate per unit volume  
$\sim\rho_b\omega^3|\Delta\xi_h|^2$. Taking the peak to have width $\sim 
\lambda$, and including a factor $R^2$ for the integration over area, we recover 
equation \refnew{eq:vis-dEdt} up to a factor of order unity.

\begin{figure*}[t]
\centerline{\psfig{figure=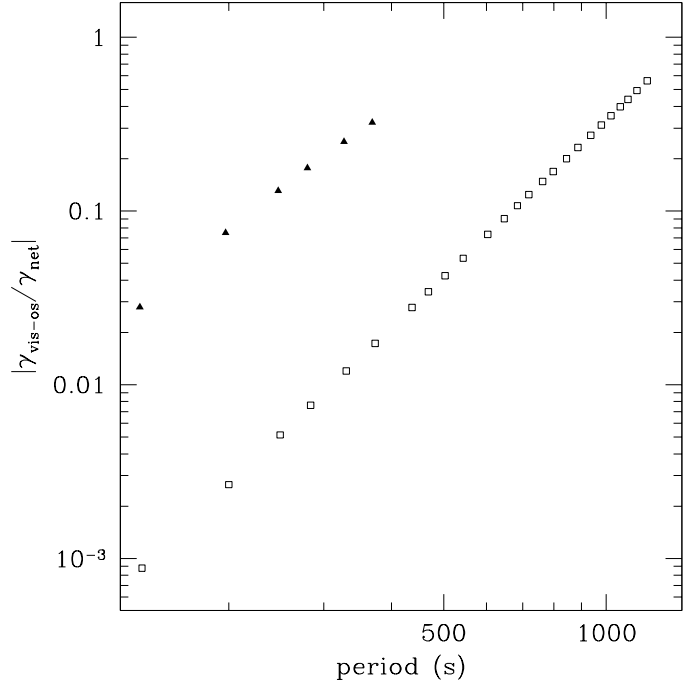,width=0.55\hsize}}
\caption[]{Ratio of the rate of linear turbulent damping in the overshoot 
region to the rate of radiative damping for two white dwarf models with
$T_{\rm eff} = 12,400 \K$ (solid triangles) and $12,000 \K$ (open
squares), plotted against mode period for $\ell = 1$
modes.  Here, $\lambda$ is taken to equal $z_b$ to maximize the effect
of overshoot damping.}
\label{fig:na-adia-vis}
\end{figure*}

To relate $\langle{dE/dt}\rangle$ to the damping rate due to convective 
overshoot, we make use of equations \refnew{eq:na-vis-dpxhequiv}, and 
\refnew{eq:na-vis-dels}, and retain only the (dominant) entropy perturbation
in equation \refnew{eq:na-vis-expressjump}. Adopting values and relations for 
$A$, $B$, $C$, $\tcv$ and $\vcv$ from Table 1, we arrive at
\be
\gamma_{\rm vis-os} \sim -0.1{\omega \tau_c\over 1+(\omega \tau_c)^2} 
\left({{\lambda}\over{z_\omega}}\right)L\left({{\delta p}\over{p}}\right)^2_b.
\label{eq:vis-osgammafinal}
\ee
Figure \ref{fig:na-adia-vis} displays $\gamma_{\rm vis-os}/\gamma_{\rm
rad}$ for g-modes in stars of $T_{\rm eff} = 12,400 \K$ and $T_{\rm
eff}=12,000
\K$.\footnote{The DA white dwarf models used in this paper are provided by P. 
Bradley. For model details, see Bradley \cite{na-vis-bradley}.}  We choose 
$\lambda/z_b=1$ in order to maximize turbulent damping in the region of 
convective overshoot.\footnote{Helioseismology constrains the entire depth of 
the region of convective overshoot in the sun to be less than 0.05
pressure scale-heights (Basu \cite{vis-basu97}).}  Even with this
extreme choice for $\lambda/z_b$, $\gamma_{\rm vis-os}$ is smaller
than $\gamma_{\rm rad}$ for all the overstable g-modes.

\section{Amplitude Limitation Due To Shear Instability
\label{sec:na-vis-shear}}

Kelvin-Helmholtz instability of the shear layer at the top of the
radiative interior is a source of nonlinear mode damping. As such, it
acts to limit the amplitudes which overstable modes can achieve. We
derive limiting amplitudes by balancing dissipation rates due to shear
instability against excitation rates due to convective driving. In this
section we deal with physical, as opposed to normalized, perturbations.

Even in carefully controlled laboratory settings, unstable shear layers are
complex structures.  Those of interest here are further complicated by
externally driven turbulence associated with penetrative convection.\footnote{We
distinguish turbulence associated with convective overshoot from that due to
instability of a g-mode's velocity shear.}  In particular, turbulent mixing
reduces the Brunt-V{\"a}is{\"a}l{\"a} frequency while turbulent viscosity lowers
the effective Reynolds number.  The former reduces stability while the latter
enhances it.  It is customary to express the nonlinear stress that maintains the
shear in terms of the velocity jump across the layer, $\Delta v$, and a
dimensionless drag coefficient, $C_D$, as (see, e.g., Landau \& Lifshitz
\cite{na-vis-landau59}, Tritton \cite{na-vis-tritton77}) \be {\cal S}={1\over 2}
C_D\rho\left(\Delta v\right)^2; \label{eq:na-vis-drag} \ee $C_D$ depends only
weakly on $\Delta v$ in turbulent shear layers.  Terrestrial experiments
indicate that $C_D$ varies logarithmically with the ratio of wall roughness to
boundary layer width, with values as small as $10^{-3}$ being characteristic of
flows over smooth walls.  We might speculate that penetrative convection makes
the upper boundary of the shear layer behave like a rough wall.  Since we have
no physical basis for assigning a reliable value to $C_D$, we treat it as a free
parameter subject to the constraint $10^{-3}\leq C_D\leq 10^{-1}$.

The power per unit area input to the shear layer is $P/A={\cal S}\Delta v$. To 
obtain the time average rate of energy dissipation in the shear layer, we 
multiply $P/A$ by $4\pi R^2$, and carry out appropriate averages over both time 
and solid angle.\footnote{The velocity jump associated with a g-mode shear layer  
has an angular dependence and varies harmonically with time.}  Next we set 
$\Delta v=\omega|\Delta \xi_h|$, where $|\Delta\xi_h|$ is evaluated using 
equation \refnew{eq:na-vis-dpxhequiv}, \refnew{eq:na-vis-dels}, and 
\refnew{eq:na-vis-expressjump} as in \S \ref{subsec:na-vis-estib}. This series 
of steps leads to
\be
\left\langle{dE_{\rm vis-nl}\over dt}\right\rangle\sim 0.1 
C_D{\omega\tau_c\over[(\omega\tau_c)^2+1]^{3/2}}{z_b\over 
k_hz_\omega^2}L\left({\delta p\over p}\right)_b^3.
\label{eq:na-vis-Edotv}
\ee

The saturation amplitude is obtained by balancing the damping due to shear 
instability by the net rate of energy gain due to convective driving plus 
radiative damping (see Paper I),
\be
\left\langle{dE_{\rm net}\over dt}\right\rangle\sim 
0.01{[(\omega\tau_c)^2-1]\over 
[(\omega\tau_c)^2+1]}L\left({\delta p\over p}\right)_b^2. 
\label{eq:na-vis-Edotcv}
\ee
This yields
\be
\left({\delta p\over p}\right)_b\sim {0.1\over 
C_D}{[(\omega\tau_c)^2+1]^{1/2}[(\omega\tau_c)^2-1]\over 
\omega\tau_c}{k_hz_\omega^2\over z_b}.
\label{eq:na-vis-dpnl}
\ee

Amplitudes of overstable modes limited by shear instability exhibit the 
following patterns. In a given star, $(\delta p/p)_b$ declines sharply with 
increasing period. For a fixed mode, the value of $(\delta p/p)_b$ rises
as the host star cools. The fractional flux perturbation at the base of
the convection zone is about $2(\delta p/p)_b$. The flux perturbation at
the photosphere is smaller by the visibility reduction factor 
$1/[1+(\omega\tau_c)^2]^{1/2}$ (see Paper I). Realistic estimates of mode 
amplitudes must await consideration of additional amplitude limiting processes 
and the calculation of nonadiabatic growth rates.

Is it safe to assume, as we have been doing, that the shear layer is
turbulent?  Necessary conditions are that the unperturbed shear layer
has both local shear large compared to $\omega$ and local Reynolds
number large compared to unity.  These conditions can be expressed by
the dimensionless relations $|d\xi_h/dz|\gg 1$ and
$Re=(\omega|\xi_h|^2)/(\nu|d\xi_h/dz|)\gg 1$. Based on the discussion
in \S \ref{subsec:na-vis-estib}, we know that the left hand side of
each inequality attains a peak value $\sim |\Delta\xi_h|/\lambda$ near
to where $\nu\sim \omega\lambda^2$ (see
Fig. \ref{fig:na-vis-plate2}). Thus the assumption of turbulence for
low $\ell$ modes requires $(\delta p/p)_b\gg
(\lambda/z_b)(z_\omega/R)$. The right hand side of this inequality is
of order $z_b/R\approx 10^{-4}$ for lowest frequency modes detected in
stars near the red edge of the instability strip.

\section{Summary
\label{sec:na-vis-resimp}}

Brickhill (\cite{vis-brick90}) discussed the manner in which turbulent
convection affects g-modes in ZZ Ceti stars. He recognized that
turbulent viscosity forces the horizontal velocity to be nearly
independent of depth inside the convection zone.  Moreover, he deduced
that turbulent damping inside the convection zone is negligible
whereas that in the region of convective overshoot might stabilize
long period modes near the red edge of the instability strip. Our
investigation supports Brickhill's conclusions, although it suggests
that turbulent damping by penetrative convection is, at best, of minor
significance.

We extend the investigation of turbulent dissipation to include that
due to instability in the shear layer at the top of the radiative
interior.  Turbulence generated by Kelvin-Helmholtz instability
depends upon the velocity shear, so the rate of energy dissipation
depends nonlinearly on the mode energy; $dE/dt\propto E^{3/2}$. Thus
shear instability provides an amplitude saturation mechanism for
overstable modes.  
 
The applicability of results from this paper to ZZ Cetis must be viewed with
caution. The quasiadiabatic approximation, while useful, is of limited
validity. Fully nonadiabatic results will be presented in Wu \& Goldreich 
(\cite{vis-paper3}).

\begin{acknowledgements}
We are indebted to Bradley for supplying us with models of DA white dwarfs. 
Financial support for this research was provided by NSF grant 94-14232.  
\end{acknowledgements}

\end{document}